\documentclass[conference]{IEEEtran}
\IEEEoverridecommandlockouts
\usepackage{cite}
\usepackage{url}
\usepackage{amsmath,amssymb,amsfonts}
\usepackage{array}
\usepackage{blkarray} 
\usepackage{makecell} 
\usepackage{algorithmic}
\usepackage{graphicx}
\usepackage[caption=false]{subfig}
\usepackage{textcomp}
\usepackage{xcolor}
\def\BibTeX{{\rm B\kern-.05em{\sc i\kern-.025em b}\kern-.08em
    T\kern-.1667em\lower.7ex\hbox{E}\kern-.125emX}}

\begin{document}

\title{Error-Mitigated Quantum Routing on Noisy  Devices
}

\author{
\IEEEauthorblockN{Wenbo Shi and Robert Malaney}
\IEEEauthorblockA{
\textit{The University of New South Wales},
Sydney, NSW, Australia. \\
}
}

\maketitle
 
\begin{abstract}


With sub-threshold quantum error correction on quantum hardware still out of reach, quantum error mitigation methods are currently deemed an attractive option for implementing certain applications on near-term noisy quantum devices. One such application is quantum routing - the ability to map an incoming quantum signal into a superposition of paths. In this work, we use a 7-qubit IBM quantum device to experimentally deploy two promising quantum error mitigation methods, Zero-Noise Extrapolation (ZNE) and Probabilistic Error Cancellation (PEC), in the context of quantum routing. Importantly, beyond investigating the improved performance of quantum routing  via ZNE and PEC separately, we also investigate the routing performance provided by the concatenation of these two error-mitigation methods. Our experimental results demonstrate that such concatenation leads a very significant performance improvement relative to implementation with no error mitigation. Indeed, an almost perfect performance in terms of fidelity of the output entangled paths is found. These new results reveal that with concatenated quantum error-mitigation embedded, useful quantum routing becomes feasible on current devices without the need for quantum error correction - opening up a potential implementation pathway to other applications that utilize a superposition of communication links.


\end{abstract}


\section{Introduction}


Quantum routers\footnote{The concept of the quantum router in this work should not be confused with classical-routing decisions for entanglement distribution \cite{pant2019routing}, but rather a quantum-only phenomenon where an input signal is routed into a coherent superposition of multiple-output communication paths.} are devices that can route quantum signals in superposition over multiple paths and are considered important elements for increasing functionality in quantum networks~\cite{network2010Duan, internet2018}, for the implementation of quantum-enabled  memory access~\cite{giovannetti2008quantum}, for the delivery of superposed quantum error mitigation~\cite{miguelramiro2023sqem}, and in assisting quantum machine learning~\cite{nawaz2019quantum}.
A quantum router has to satisfy six key requirements~\cite{Lemr2013resource, yuan2015experimental}: 
(i)~Both signal and control information are stored in qubits.
(ii)~The signal information is preserved after the routing process.
(iii)~The signal should be routed to a coherent superposition of both output paths.
(iv) No post-selection for signal qubits.
(v)~Only one control qubit is utilized for routing one signal qubit.
(vi)~Entanglement between control and signal qubits is generated after the quantum routing.

The work of~\cite{chang2013experimental} provided the first proof-of-principle experimental demonstration of a probabilistic quantum router using entangled photons.
However, this quantum router collapsed the signal information, therefore not meeting the requirement (ii).
A proposal for an all-linear-optical quantum router with $1/4$ success probability was given in~\cite{LEMR2013282}, and an implementation of this proposal was demonstrated in~\cite{Bartkiewicz2018implementation}.
Nevertheless, two control qubits were required for routing one signal qubit in this demonstration, therefore not meeting  the requirement (v).
In~\cite{Lemr2013resource}, an all-linear-optical quantum router meeting the requirements (i)-(v) was experimentally demonstrated, although the  success probability was only $1/8$.
The work of~\cite{yuan2015experimental} proposed an optical quantum router meeting all six requirements.
This router was deterministic but challenging for experimental implementation - only a probabilistic version of the router was illustrated.
In~\cite{Bikash2019Designing}, a deterministic quantum router using superconducting qubits was designed and experimentally demonstrated.


Near-term quantum devices, the so called Noisy Intermediate-Scale Quantum (NISQ) devices are now widely available~\cite{Preskill2018quantumcomputingin}.
The high error rates of these current quantum devices usually prevent the realization of quantum applications, and straightforward implementation of quantum error correction is generally not possible on them. As such, \emph{quantum error mitigation} provides a potential alternate pathway to application deployment - a possibility that has recently attracted widespread attention.
Quantum error mitigation methods aim to reduce the effects (e.g. magnitude) of system errors rather than completely eliminate them~\cite{Lowe2021Unified}.
Two well-known quantum error mitigation methods are Zero-Noise Extrapolation (ZNE)~\cite{temme2017zne, Kandala2019zne, Li2017zne} and Probabilistic Error Cancellation (PEC)~\cite{Mari2021NEPEC, reviewQEM2022}, which utilize classical post-processing approaches to mitigate errors~\cite{Bultrini2022Simple}.

\begin{figure*}[tb]
    \centering
    \includegraphics[width=\textwidth]{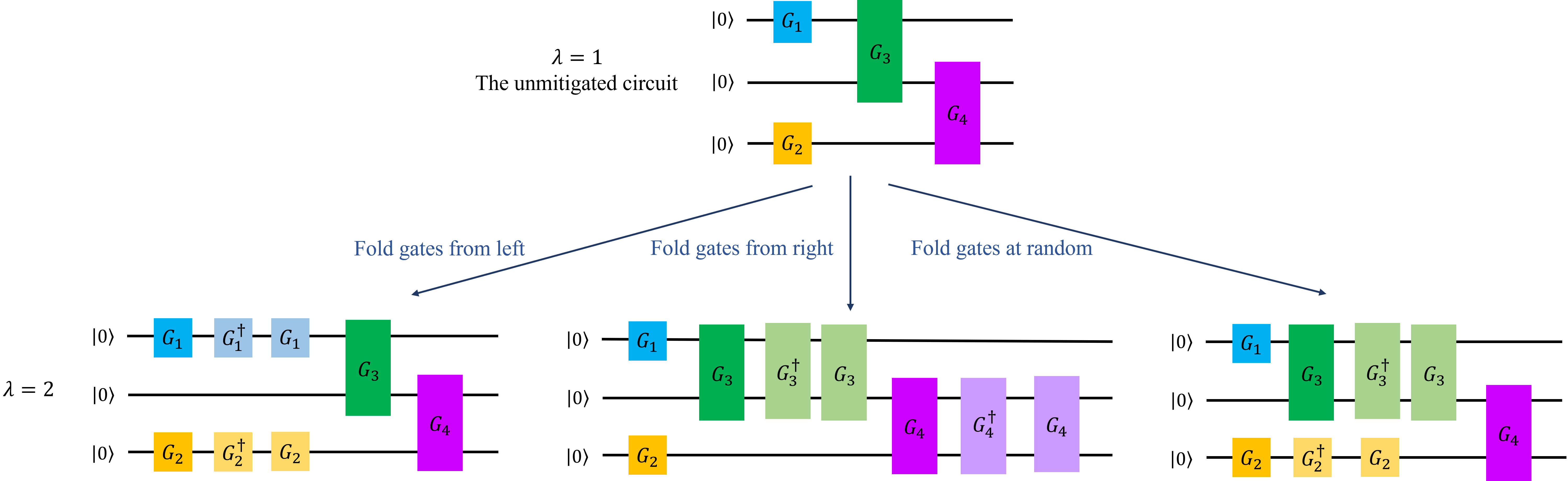}
    \caption{Three methods of local folding gates in ZNE. The value of $\lambda$ decides how many gates should be inserted to extend the strength of the noise. The total number of gates after the insertion should be approximately $\lambda$ times the number of gates in the unmitigated circuit.}
    \label{fig:zneFold}
\end{figure*}

\begin{figure*}[t]
    \centering
    \includegraphics[width=\linewidth]{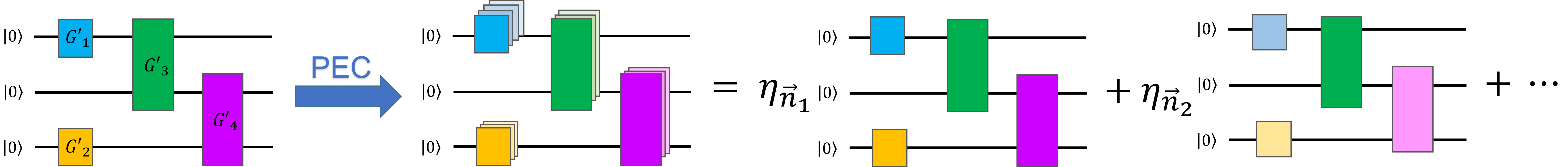}
\caption{Working principle of PEC. 
Suppose that a quantum circuit includes four noiseless unitary gates, namely $G'_1$, $G'_2$, $G'_3$, and $G'_4$.
Each unitary gate can be represented by a group of implementable but noisy gates.
Therefore, the quantum circuit can be represented by a linear combination of noisy quantum circuits with implementable gates only.
}
    \label{fig:PECprinciple}
\end{figure*}

In this work we focus on NISQ devices  manufactured by IBM - superconducting quantum devices made accessible to the research community~\cite{ibmq}. 
On such devices, our previous work~\cite{entropy2023wenbo} experimentally realized the quantum router with a quantum error-correcting code of~\cite{Laflamme1996Perfect}.
However, the results of~\cite{entropy2023wenbo} demonstrated that such error correction is, in general, ineffective on the current IBM quantum devices in the context of quantum routing.
In this work, we apply, for the first time, two quantum error mitigation techniques (ZNE and PEC) in a concatenated form to the quantum routing problem.
We shall see that these quantum error mitigation techniques significantly improve the entanglement fidelity of the quantum router - to the point that  quantum applications based on quantum routing become effective on current devices. 

Although our work focuses on the IBM devices, we believe that superconducting-based quantum routers can be generalized for use in future large-scale quantum communication networks via interface techniques which convert photons to superconducting qubits and vice versa \cite{Supercondu2018, Mirhosseini2020}. As such, we believe the new results reported here will have wider implications beyond routing within the computer hardware itself - important as that is.
The rest of this paper is organized as follows.
Section~\ref{Principle} introduces the working principles of ZNE and PEC. Section~\ref{EX} presents the quantum routing protocol as well as the experimental results. A brief discussion on the use of quantum routing in a related application is also given there.
Section~\ref{Conclusions} concludes this work.

\section{Quantum Error Mitigation} \label{Principle}
    \subsection{Overview of ZNE}


Suppose that $E$ is a Hermitian operator that has discrete eigenvalues $\{a_1, a_2,\cdots, a_b\}$ with associated eigenstates $\{ \vert a_l\rangle \text{, } l = 1,2,\cdots, b \}$, where $b$ represents the number of eigenvalues.
These eigenstates form a complete and orthonormal set such that any state $\vert \chi \rangle$ can be written as 
$\vert \chi \rangle = \sum_{l} \vert a_l \rangle\langle a_l \vert \chi \rangle$.
The probability $P$ of measuring $\vert \chi \rangle $ to be in the state $\vert a_l\rangle$ via a measurement of $E$ is 
$P = \vert \langle a_l \vert \chi \rangle \vert^2 
= \frac{A_l}{A}
\text{.}$
Here, $A$ is the number of copies of $\vert\chi\rangle$  and $A_l$ represents the number of these that  become $\vert a_l \rangle$ upon measurement of $E$.
In  most calculations presented here, the eigenstates will be the three-qubit states $\vert \varphi \rangle^{\otimes 3}$, where $\vert \varphi \rangle \in \{ \vert0\rangle, \vert1\rangle\}$.

As one of the main quantum error mitigation techniques, ZNE extrapolates to the zero-noise limit of a quantum device by scaling the strength of noise~\cite{Li2017zne, temme2017zne}.
ZNE estimates an ideal (noiseless) expectation value $\left< E \right>_{ideal} \equiv \text{Tr}\left[ E \vert \chi \rangle \langle \chi \vert \right]$ from a group of noisy measurements by extrapolating to the zero-noise case.
The $\left< E \right>_{ideal}$ denotes the expected result in a measurement, \textit{i.e.}, $\left< E \right>_{ideal} = \sum_l a_l \vert \langle a_l\vert \chi \rangle \vert^2$.
The strength of the noise is represented by a noise scale factor, $\lambda \geq 1$, and the expectation value of $E$ for that $\lambda$ is represented as $\left< E \right>_{noisy}^{\lambda}$.


Generally, three noise-scaling methods can be utilized: (i)~pulse stretching, (ii)~local folding, and (iii)~global folding.
(i)~Pulse stretching is an analog method that amplifies the noise of the quantum device by stretching microwave gate pulses to longer time intervals.
(ii)~Local folding maps unitary gates in a quantum circuit with a mapping logic of $ G \rightarrow G \left( G^\dag G \right)^{k} $, where $G$ represents an implementable gate and $k$ is a positive integer influenced by $\lambda$.
With a given $\lambda$, one could fold some or all gates in a quantum circuit with $k \geq 1$.
The total number of quantum gates of a folded circuit is approximately $\lambda$ times of the number of gates of the unmitigated circuit ($\lambda =1$ circuit).
Three methods can be utilized for inserting the folding gates locally, namely, from the left, from the right, or at random (see Fig.~\ref{fig:zneFold}).
(iii)~Global folding utilizes the same mapping logic but folds a collection of unitary gates together, \textit{i.e.}, $U\rightarrow U\left(U^\dag U\right)^{k}$, where $U$ is used to represent a circuit~\cite{giurgica2020zneFolding}.


\begin{figure*}[ht]
    \centering
    \includegraphics[width =\textwidth]{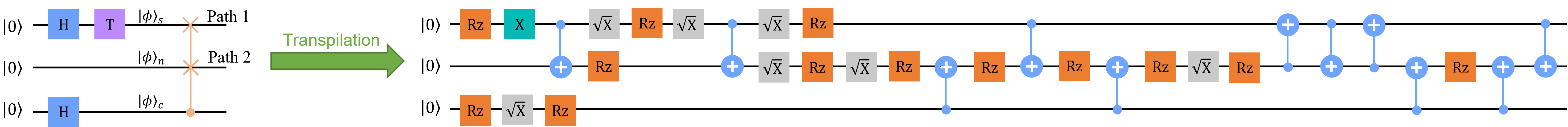}
\caption{Quantum circuits of the quantum routing protocol before and after transpilation.
In the circuits, H stands for the Hadamard gate, T is a phase gate introducing a $\pi/4$ phase, X represents the NOT gate, $\sqrt{\text{X}}$ rotates a qubit about the X-axis by $\pi/2$, and Rz rotates a qubit about the Z-axis with a given phase.
The three-qubit gate is the controlled-swap gate which realizes the quantum routing process, and the two-qubit gate is the Controlled-X (CX) gate.
}
    \label{fig:qr}
\end{figure*}

If one chooses $\lambda = \left[ \lambda_1, \lambda_2, \cdots, \lambda_j \right]$, then $j$ ancillary quantum circuits would be constructed accordingly.
After executing these ancillary circuits on the quantum device, $j$ values of $\left< E \right>_{noisy}^{\lambda}$ can be collected.
Extrapolation is implemented as a post-processing method acting on the collected $\left< E \right>_{noisy}^{\lambda}$.
Multiple extrapolation methods, such as linear extrapolation, polynomial extrapolation, and exponential extrapolation, can be considered.
These methods fit the curve plotted by $\left< E \right>_{noisy}^{\lambda}$ as a function of $\lambda$.
Based on the fitted curve, $\left< E \right>_{ideal}$ can be found via extrapolation to $\lambda = 0$.

    \subsection{Overview of PEC}

Different from ZNE,  PEC attempts to represent noiseless gates (non-implementable) with noisy  gates (implementable). The working principle of PEC is illustrated in Fig.~\ref{fig:PECprinciple}.
Assume that a unitary circuit $U'$ is applied to $m$ qubits initially prepared as $\vert 0\rangle^{\otimes m}$, and $U'$ is composed of $T$ noiseless gates $G'_{t}$, where $t \in \{1,2,\cdots, T\}$.
Suppose that $G'_{t}$ also indicates the specific qubits that are to be applied.
This noiseless circuit, $U'$, can be expressed as $U' = G'_T\cdots G'_2G'_1$, which is a noiseless version of $U$.



The first core step of PEC is the representation of each $G'_{t}$, by a set of noisy, but implementable, gates $\{ G_{t,n_t} \}$ in the form of
\begin{equation} \label{G}
\begin{split}
G'_{t} &= \sum_{n_t} \eta_{t,n_t} G_{t,n_t} \\
& = \eta_{t,1_t} G_{t,1_t} + \eta_{t,2_t} G_{t,2_t} + \cdots + \eta_{t,N_t} G_{t,N_t}
\text{,}
\end{split}
\end{equation}
where 
$\eta_{t,n_t}$ are real coefficients and satisfy the trace-preserving condition, \textit{i.e.}, $\sum_{n_t} \eta_{t,n_t} =1$.
The number of gates, $n_t$, required to represent $G'_{t}$ at any time is a variable between 1 and $N_t$.
Note that $n_t$ varies gate by gate and is dependent on $t$.
The method of deriving the noisy representation of a specific noiseless gate in our experiments is detailed in Section \ref{EXsetup}.

The second core step of PEC is the estimation of $\left< E \right>_{ideal}$ by  sampling  from the noisy representations via a Monte Carlo average~\cite{temme2017zne, PracticalEndo2018}.
If the noisy representation of each noiseless gate in $U'$ is known, then $\left< E \right>_{ideal}$ can be estimated as a Monte Carlo average over different noisy circuits, where each noisy circuit is sampled according to the noisy representations.
In principle, with full tomographic knowledge of $\{ G_{t,n_t} \}$ and sufficiently large number of samples, one could cancel all hardware noise~\cite{Mitiq}.
Replacing each $G'_{t}$ in $U'$ with its noisy representation, $\left< E \right>_{ideal}$ can be expressed as
\begin{equation} \label{E}
\left< E\right>_{ideal} = 
\text{Tr} \left[ U'(\rho_0)  E\right]
= \sum_{\vec{n}} \eta_{\Vec{n}} \left< E_{\vec{n}} \right>_{noisy}
\text{,}
\end{equation}
where $\rho_0$ is the initial state, $\vert 0\rangle^{\otimes m}$, of $U'$,
\begin{equation}
\eta_{\vec{n}} := \prod_{t=1}^{T} \eta_{t,n_t}
\text{,  }
\mathbb{U}_{\vec{n}} := \prod_{t=1}^{T} G_{t,n_t}
\text{, and}
\end{equation}
\begin{equation}
\left< E_{\vec{n}} \right>_{noisy} :=
\text{Tr} \left[ \mathbb{U}_{\vec{n}} \left(\rho_0 \right) E\right]
\text{, such that}
\end{equation}
\begin{equation}
U' = \sum_{\vec{n}} \eta_{\vec{n}} \mathbb{U}_{\vec{n}} 
= \prod_{t=1}^{T} \left( \sum_{n_t} \eta_{t,n_t} G_{t,n_t} \right)
\text{.}
\end{equation}
Again, $\eta_{\vec{n}}$ are real coefficients, satisfying $\sum_{\vec{n}} \eta_{\vec{n}} =1$~\cite{Mari2021NEPEC}.
Note that $\mathbb{U}_{\vec{n}}$ represents the unitary of a noisy quantum circuit that is sampled from the noisy representations of the noiseless gates and is regarded as an ancillary circuit of PEC.
The number of $\mathbb{U}_{\vec{n}}$ is the number of samples, and we denote this number as $s$.
All $\left< E_{\vec{n}} \right>_{noisy}$ can be obtained from the NISQ devices since the $\mathbb{U}_{\vec{n}}$ only require implementable gates.
With sufficient and a suitable linear combination of $\left< E_{\vec{n}} \right>_{noisy}$, one can obtain an unbiased estimate of $\left< E\right>_{ideal}$.

\section{Experiments} \label{EX}

    \subsection{Quantum Routing with Error Mitigation}

We now apply the mitigation methods to  quantum routing realized on the  \textit{ibmq\_jakarta}  superconducting quantum device - a device which  has seven physical qubits with a horizontal H shape~\cite{ibmq}.
We use the quantum routing protocol utilized in~\cite{entropy2023wenbo} as the application to benchmark the performance of the mitigation methods.
This routing protocol requires three qubits: a signal qubit $\vert \phi_s\rangle$, a control qubit $\vert \phi_c\rangle$, and a null qubit $\vert \phi_n\rangle = \vert0\rangle_n$, which is an ancillary qubit of the quantum router.
We define 
$\vert\phi_s\rangle = 
\cos(\pi /4)\vert0\rangle_s + e^{i\pi/4} \sin(\pi /4) \vert1\rangle_s$ representing the transmitted signal information, and note $\vert\phi_s\rangle$ can be prepared by sequentially implementing the Hadamard gate and the T phase gate (which introduces a $\pi/4$ phase).
We prepare $\vert\phi_c\rangle= 
(\vert0\rangle_c+\vert1\rangle_c)/\sqrt{2}$ via the Hadamard gate, where $\vert\phi_c\rangle$ stands for the control information directing the path of $\vert \phi_s\rangle$.
The signal qubit is injected into the quantum router via path~$1$, and the null qubit is initially prepared at path~$2$.
The input of the quantum router is thus given by $\vert\Phi\rangle = \vert\phi_s\rangle_1 \vert\phi_n\rangle_2 \vert\phi_c\rangle$, where the subscripts $1$ and $2$ represent the paths~$1$ and $2$, respectively.
The quantum router routes the signal qubit to the paths~$1$ and $2$ simultaneously since the control qubit is in a superposition state.
Therefore, the output of the quantum router is 
$\vert\Phi\rangle_f =
\left(
\vert\phi_s\rangle_1\vert\phi_n\rangle_2 \vert0\rangle_c 
+ \vert\phi_n\rangle_1 \vert\phi_s\rangle_2 \vert1\rangle_c 
\right) /\sqrt{2}$, which is an entanglement between the control qubit and the two paths.

We build a quantum circuit to realize the quantum routing protocol, and then execute this quantum circuit  via a transpiled version of the circuit - see Fig.~\ref{fig:qr}.
The quantum gates in the transpiled circuit can be implemented directly on the quantum device.
The construction of the quantum circuit and the implementation of the transpilation are realized via IBM's open-source software development kit---the Quantum Information Science toolKit (Qiskit)~\cite{aleksandrowicz2019qiskit}.
We choose three physically connected qubits with relatively low quantum gate error rates  to act as the three qubits of the transpiled circuit.
Note that the quantum gate error rates may change over time - the \textit{ibmq\_jakarta} is calibrated daily.

    \subsection{Experimental Setups} \label{EXsetup}


The two quantum error mitigation methods, ZNE and PEC, mainly focus on quantum gate errors, decoherence errors, and cross-talk errors - they cannot mitigate measurement errors.
Measurement errors mistakenly read a qubit in the $\vert 0\rangle$ state as the $\vert 1\rangle$ state, and vice versa.
We choose a measurement error mitigation protocol~\cite{MeasurementQiskit} provided by Qiskit to reduce measurement errors.
This protocol  requires ancillary quantum circuits, which we call calibration circuits.
One needs $2^m$ calibration circuits to construct a $2^m\times2^m$ calibration matrix $M$, where $m$ is the number of qubits that are measured.
Each calibration circuit prepares the $m$ qubits to one of $2^m$ Z-basis states before using the Z-basis measurements to measure them, where the Z-basis states are $\vert \varphi \rangle^{\otimes m}$ and $\vert \varphi \rangle \in \{ \vert0\rangle, \vert1\rangle\}$.
The measurement results of the calibration circuits determine $M$, and  $M^{-1}$  is then applied to the experimental results to eliminate measurement errors. Typically, fidelity improvements of order $10\%$ are found via elimination of measurement errors.
Henceforth by the term ``unmitigated" we will mean without any ZNE or PEC included in the results  - measurement error mitigation is by default included in \emph{all} results we show here. The term ``error mitigation" will henceforth refer only to ZNE and/or PEC.

To investigate the experimental performance of the quantum routing protocol with quantum error mitigation, we utilize an open-source package named Mitiq~\cite{Mitiq} to implement ZNE and PEC.
Although Qiskit recently released built-in functions for employing quantum error mitigation methods (twirled readout error extinction, ZNE, and PEC)~\cite{qiskitQEM}, one cannot implement these methods step by step and obtain detailed data.
Another reason why we choose Mitiq instead of Qiskit's built-in functions is that we can concatenate multiple mitigation methods through Mitiq.
For implementing ZNE, we set the noise scale factor $\lambda = \left[ 1,3,5,7,9,11,13 \right]$, and we choose local folding at random to scale the noise.
Seven noise-scaled circuits are generated based on the seven values of $\lambda$, and each noise-scaled circuit is executed $A = 100,000$ times (this number of executions applied to all experiments).
We choose polynomial extrapolation with order~$2$ to extrapolate to $\left< E \right>_{ideal}$.

In principle, we need to acquire full tomographic knowledge of the quantum gates in the transpiled circuit to implement PEC.
However, to simplify our experiments, we make two assumptions: 
(i) We assume we can neglect single-qubit gate errors and only focus on two-qubit gate errors since the two-qubit gate error rates are an order of magnitude higher.
(ii) We assume that the two-qubit gates are followed by a global depolarizing noise.
The transpiled circuit only has one type of two-qubit gate, the CX gate.
Based on assumption (ii), we have
\begin{equation} \label{eq:Gcx}
G_{noisy}^{CX} = \mathcal{D}\circ \mathcal{P} \circ G_{ideal}^{CX}
\text{,}
\end{equation}
where $G_{noisy}^{CX}$ is the implementable CX gate that we assumed, $G_{ideal}^{CX}$ is the noiseless CX gate, and $\mathcal{P} \in \{I,X,Y,Z\}^{\otimes2}$ is a Pauli trace-preserving completely positive map.
Note that 
\begin{equation} \label{D}
\mathcal{D} (\hat{\rho})= (1-\epsilon) \hat{\rho} + I \epsilon/ 4  ,
\end{equation}
represents the two-qubit depolarizing channel, where $\hat{\rho}$ stands for the input state of this channel and $\epsilon$ is the noise level of the CX gate~\cite{temme2017zne}.
For each CX gate in the transpiled circuit, we acquire the associated CX gate error $\epsilon$, which varies over time, from the calibration data reported in~\cite{ibmq}.
Based on Eqs.~\eqref{eq:Gcx} and \eqref{D}, and with known $\epsilon$, $G_{ideal}^{CX}$ can be represented by a group of noisy gates in the form of Eq.~\eqref{G}.
Then, using noisy representation of the noiseless CX gate, we conduct a Monte Carlo sampling process via Mitiq with $s = 20$ ($20$ ancillary circuits are generated for PEC).
From the many executions, the measurement results of the ancillary circuits are collected to calculate $\left< E \right>_{ideal}$.

To concatenate ZNE and PEC, we first fold gates (from the left) of the transpiled circuit with $\lambda = \left[ 1,3,5,7,9 \right]$, generating five noise-scaled circuits.
We then apply PEC to each noise-scaled circuit with $s=20$, which means that we represent each CX gate in the noise-scaled circuits by its noisy representation that we derived before.
From the many executions, the error-mitigated data of the noise-scaled circuits are collected, the data will be utilized for extrapolating to $\left< E \right>_{ideal}$, again by polynomial extrapolation with order~$2$.

\begin{figure}[t]
\centering
\includegraphics[width = \linewidth]{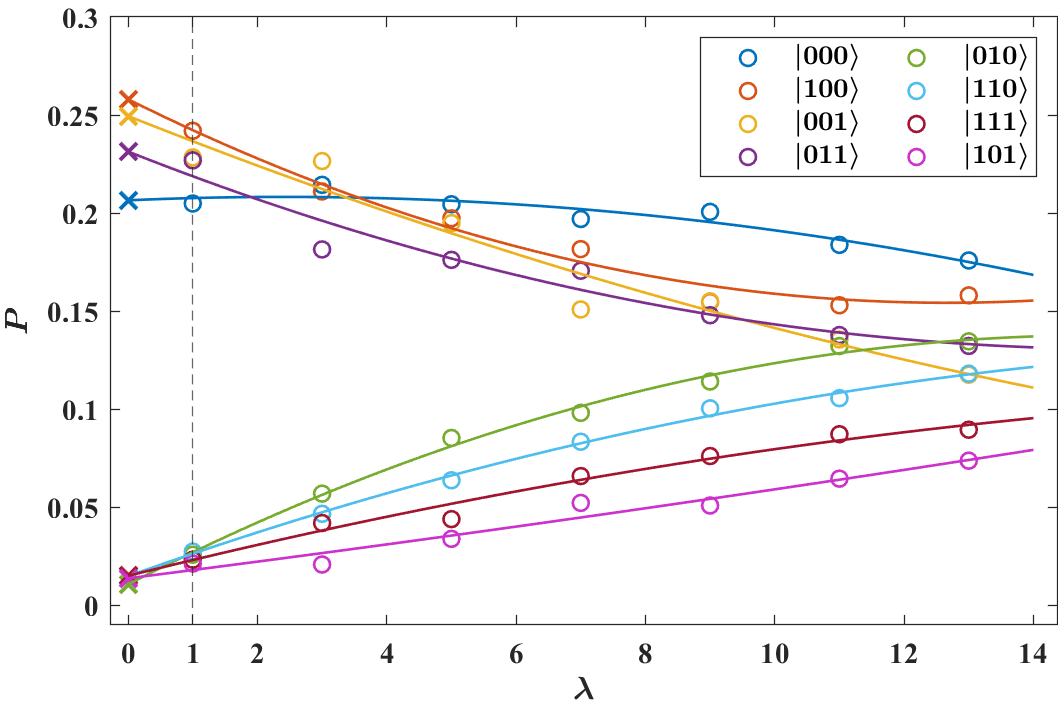}
\caption{$P$ as a function of $\lambda$ using ZNE.
The circles represent $P$ obtained from the quantum device, and the circles plotted on the vertical dashed lines indicate the unmitigated results of $P$.
The solid lines are polynomial fitted curves with order~$2$, and the cross markers stand for the corresponding mitigated results of $P$.
}
\label{figs: zne}
\end{figure}

\begin{figure}[t]
    \centering
    \includegraphics[width =\linewidth]{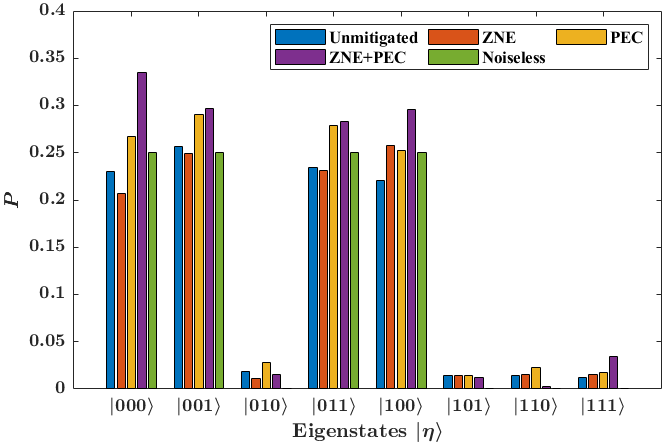}
\caption{Results using ZNE or/and PEC obtained from the quantum device.
The green bars are noiseless results plotted for reference, and the missing green bars indicate that the noiseless results should be~$0$.
}
    \label{fig:barPEC}
\end{figure}

    \subsection{Experimental Results}



We define that $\rho = \vert\Phi\rangle_f \langle\Phi\vert$ is the ``theoretical'' density matrix of the quantum router's output, \textit{i.e.}, the final state of the transpiled circuit executed on a noiseless quantum device.
We define $\rho'$ as the ``experimental'' density matrix of the quantum router's output obtained from the quantum device.
Instead of using the expectation values, 
we utilize $P$ as one of our performance metrics.
In this quantum routing experiment, $P$ is the probability of measuring $\rho'$ to be in one of the eigenstates $\vert \eta \rangle = \vert \varphi \rangle^{\otimes 3}$ with the measurement operator $Z\otimes Z \otimes Z$.

The values of $P$ as a function of $\lambda$ using ZNE are demonstrated in Fig.~\ref{figs: zne}.
In the noiseless situation, the probability of observing one of the $\vert000\rangle$, $\vert001\rangle$, $\vert011\rangle$, and $\vert100\rangle$ states is~$0.25$, and the probability of observing one of the
remaining states is~$0$.
We see that the noise grows with increasing $\lambda$, and the extrapolated results are closer to (or even the same as) the noiseless results, indicating that ZNE is an effective method to mitigate quantum errors.
The unmitigated and mitigated $P$ using different error mitigation methods are  demonstrated in Fig.~\ref{fig:barPEC}.
One can observe that $P$ are closer to the noiseless results after introducing an error mitigation method.
However, some over-correction can be noticed, especially when the concatenation of ZNE and PEC is applied.

Beyond $P$, we  choose entanglement fidelity $F$ as our main performance metric, where $F$ is expressed as 
$F= \left( \text{Tr} \sqrt{ \sqrt{\rho} \,\, \rho' \sqrt{\rho}} \right) ^2$.
Note that $\rho'$ is reconstructed by quantum state tomography, which requires at least $3^3=27$ copies of the transpiled circuits (since there are three qubits in the circuit) to apply~$27$ measurements of operators $\{X\otimes X\otimes X, X \otimes X \otimes Y, \cdots, Z \otimes Z \otimes Z\}$, which are tensor products of~$3$ Pauli operators.
For each transpiled circuit with a distinct measurement operator, we apply ZNE or/and PEC to improve its corresponding values of $P$.
One can reconstruct $\rho'$ with $P$ and the corresponding operators via the quantum state tomography.

\begin{figure}[tb]
    \centering
    \includegraphics[width =\linewidth]{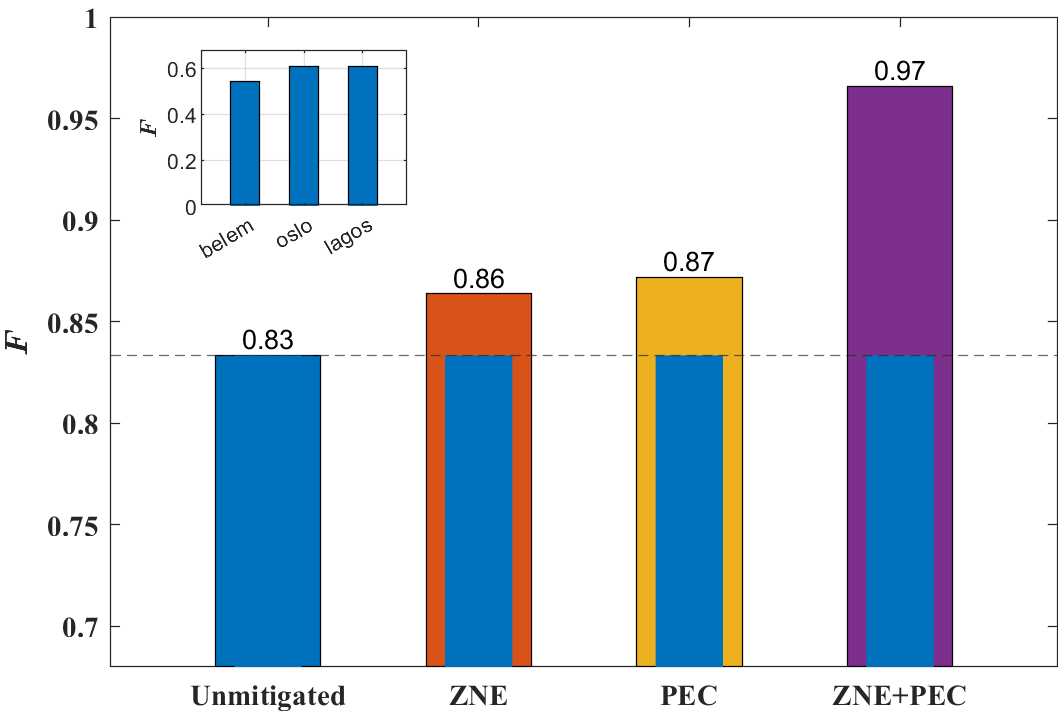}
\caption{$F$ of the quantum router with and without quantum error mitigation techniques.
The dashed horizontal line indicates the fidelity with measurement error mitigation only.
The inset figure indicates the unmitigated result of the quantum routing protocol conducted on other quantum devices, namely \textit{ibmq\_belem}, \textit{ibm\_oslo}, and \textit{ibm\_lagos}.
One can observe that these machines obtained lower values of $F$ compared to the one of the \textit{ibmq\_jakarta}, such that we take the unmitigated value $F=0.83$ as the baseline for comparing the mitigated results.
}
    \label{fig:barFid}
\end{figure}

The values of $F$ determined with different error mitigation methods are plotted in the main part of Fig.~\ref{fig:barFid}. For comparison, in the inset of Fig.~\ref{fig:barFid}
 we also demonstrate the unmitigated  $F$ values determined from other IBM's quantum devices, namely, the \textit{ibmq\_belem}, \textit{ibm\_oslo}, and \textit{ibm\_lagos}. These inset values show that the machine we use here, the \textit{ibm\_jakarta}, offers the best performance in terms of unmitigated results and therefore forms the best starting point to apply mitigation methods to. Reverting back to the main part of the figure, we can see the
fidelity results for singular use of ZNE and PEC are similar, showing a minor improvement in $F$ compared to the unmitigated result obtained from the \textit{ibm\_jakarta}.
However, interestingly the concatenation method demonstrates $F\approx 1$, representing an almost-perfect performance.
We should note this improvement does come at a cost - the concatenation method has a higher resource requirement and demands a longer execution time compared to singular use of ZNE or PEC.
\emph{The almost-unity fidelity outcome of Fig.~\ref{fig:barFid} represents the main result of this work.}

\subsection{QRAM with Error Mitigation}

\begin{figure}[t]
    \centering
    \includegraphics[width = 0.9\linewidth]{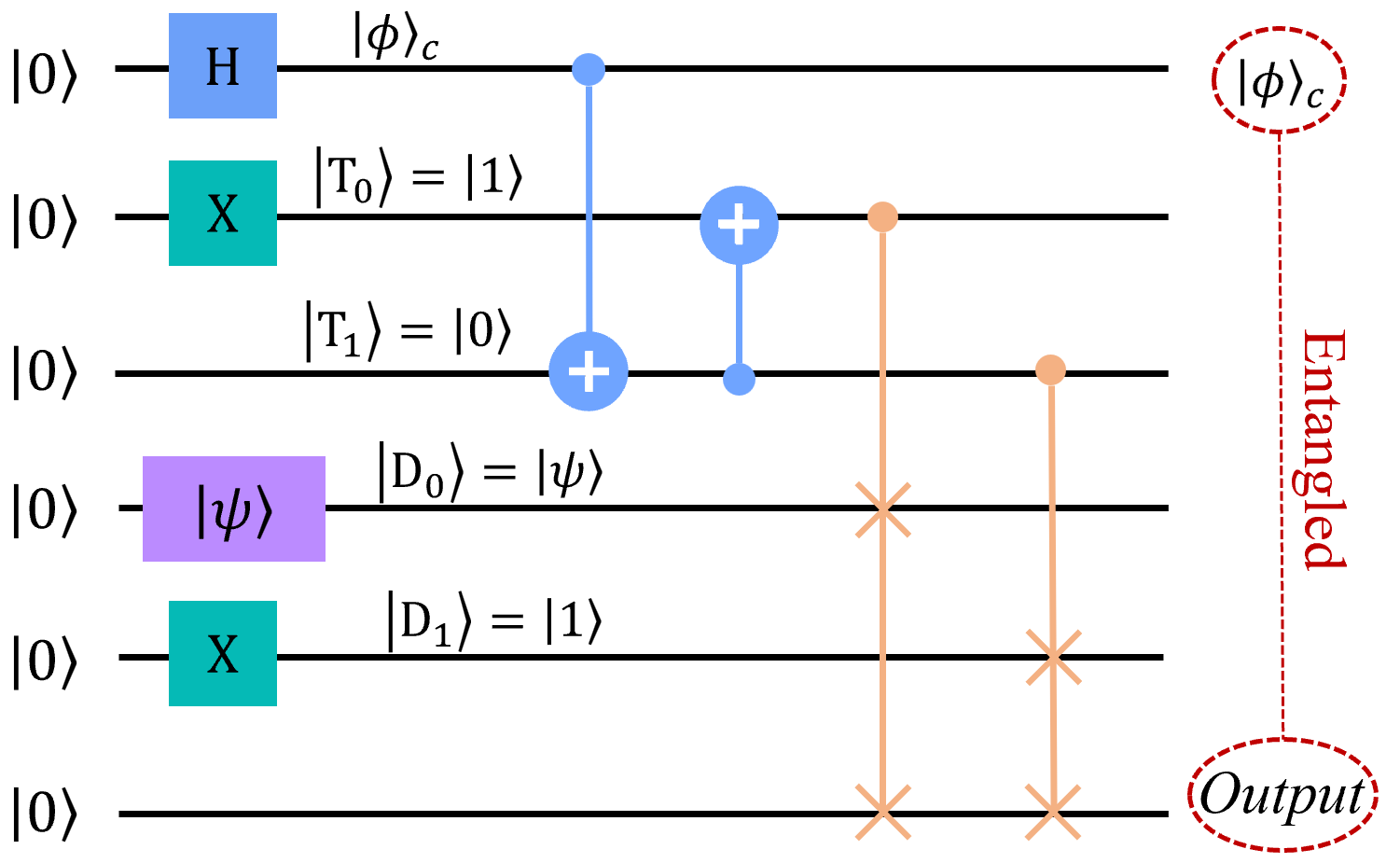}
\caption{An example of a QRAM circuit.
The purple gate indicates the preparation of $\vert \psi\rangle$.}
    \label{fig:qram}
\end{figure}

It is perhaps useful to close this work with a short discussion on another application that uses quantum-routing-like functionality.
Quantum Random Access Memory (QRAM) is an analogy of classical RAM - a critical element of a classical computing architecture.
Different from classical RAM, however, QRAM allows one to query a superposition of memory addresses that store either classical or quantum information~\cite{qramFault2020Matteo}.
A QRAM query is given by
\begin{equation}
\sum_d \alpha_d \vert d\rangle\vert0\rangle
\xrightarrow{\text{QRAM}}
\vert\Psi\rangle_{f} = 
\sum_d \alpha_d\vert d\rangle \vert D_d\rangle
\text{,}
\end{equation}
where $\sum_d \alpha_d \vert d\rangle$ is a superposition of queried addresses and $\vert D_d\rangle$ stands for the quantum or classical data stored in address $\vert d \rangle$. This illustrates the similarities of QRAM with quantum routing.
One of the most efficient QRAM schemes is the bucket brigade scheme, which employs a binary-tree-based query architecture~\cite{Arunachalam2015on}.
Based on this scheme, we construct a bucket-brigade style QRAM circuit, as shown in Fig.~\ref{fig:qram}.
The qubit $\vert\phi\rangle_c$ in this QRAM circuit represents an address qubit, including the query information (addresses that one wants to query).
The qubits $\vert T_0\rangle$ and $\vert T_1\rangle$ stand for the binary tree and $\vert D_0\rangle$ and $\vert D_1\rangle$ are data stored at the memory cells.
We define that $\vert D_0\rangle$ is a random qubit storing quantum information and $\vert D_1\rangle = \vert 1\rangle$ stores classical information. 
The last qubit in the QRAM circuit is regarded as the output of the QRAM and stores all of the data queried by the address qubit.
For this QRAM circuit, 
$\vert\Psi\rangle_{f} = \left( \vert0\rangle_c \vert D_0 \rangle_{out} 
+ \vert1\rangle_c \vert D_1 \rangle_{out} \right)/\sqrt{2}$, since we define the address qubit as $\vert \phi\rangle_c$.

We again chose the entanglement fidelity $F$ as our performance metric but with the theoretical density matrix $\rho = \vert\Psi\rangle_{f} \langle\Psi \vert$ and the experimental density matrix reconstructed by applying quantum state tomography on the first (count from the top) and the last qubits of the QRAM circuit illustrated.
However, from our QRAM experiments we find the maximum increase in fidelity due to error mitigation is only $5\%$ (maximum fidelity found being 0.76). This phenomenon is caused by the complexity of the transpiled circuits, relative to the transpiled  quantum routing application discussed earlier.
In particular,
one can observe that the QRAM circuit includes two controlled-swap gates (see Fig.~\ref{fig:qram}), while the quantum routing circuit requires one such gate.
The increase in the number of controlled-swap gates leads to a longer time and a higher complexity for the execution of QRAM.
The longer execution time amplifies the decoherence errors, and the increase of the number of gates in the execution accumulates the gate errors.
Future work for implementing QRAM on the NISQ devices should focus on the improvement of the quantum error mitigation methods designed for complicated quantum circuits.

\section{Conclusions} \label{Conclusions}

In this work, we experimentally tested the performance of two quantum error mitigation methods, ZNE and PEC, implemented in the context of a quantum routing protocol.
Our results indicated that utilizing ZNE or PEC helped improve the performance of the near-term quantum devices.
More importantly, we found concatenating ZNE and PEC impressively increased the entanglement fidelity of the quantum router to effectively one.
This result reveals the capacity of concatenating quantum error mitigation methods as applied to NISQ devices. 
Although quantum error mitigation methods require additional executions of ancillary quantum circuits, our results show the critical role such methods can have for applications run on current NISQ devices.
More specifically, our results provide an overview of what can be anticipated for an error-mitigated quantum router in practice - illustrating that full-blown quantum error correction processes need not be implemented on  current devices for this important quantum application. 



\bibliographystyle{IEEEtran}
\bibliography{IEEEabrv,References}

\end{document}